\documentclass{elsart4-1}

\usepackage{graphicx}

\usepackage{amssymb}
\usepackage[latin1]{inputenc}

\usepackage[english,francais]{babel}

\newtheorem{e-proposition}[theorem]{Proposition}

\newtheorem{e-definition}[theorem]{Definition\rm}

\renewcommand{\theequation}{\arabic{equation}}
\newcommand{\be}{\begin{equation}}
\newcommand{\ee}{\end{equation}}
\def\og{\leavevmode\raise.3ex\hbox{$\scriptscriptstyle\langle\!\langle$~}}
\def\fg{\leavevmode\raise.3ex\hbox{~$\!\scriptscriptstyle\,\rangle\!\rangle$}}

\begin{document}

\centerline{Astrophysics}
\begin{frontmatter}


\selectlanguage{english}
\title{Origin of very high and ultra high energy cosmic rays}


\selectlanguage{english}
\author[blasi1,blasi2]{Pasquale Blasi},
\ead{blasi@arcetri.astro.it}

\address[blasi1]{INAF/Osservatorio Astrofisico di Arcetri, Largo E. Fermi, 5 - 50125 Firenze, Italy}
\address[blasi2]{Gran Sasso Science Institute (INFN), viale F. Crispi 7, 67100 L' Aquila, Italy}

\medskip
\begin{center}
{\small Received *; accepted after revision +}
\end{center}

\begin{abstract}
While there is some level of consensus on a Galactic origin of cosmic rays up to the knee ($E_{k}\sim 3\times 10^{15}$ eV) and on an extragalactic origin of cosmic rays with energy above $\sim 10^{19}$ eV, the debate on the genesis of cosmic rays in the intermediate energy region has received much less attention, mainly because of the ambiguity intrinsic in defining such a region. The energy range between $10^{17}$ eV and $\sim 10^{19}$ eV is likely to be the place where the transition from Galactic to extragalactic cosmic rays takes place. Hence the origin of these particles, though being of the highest importance from the physics point of view, it is also one of the most difficult aspects to investigate. Here I will illustrate some ideas concerning the sites of acceleration of these particles and the questions that their investigation may help answer, including the origin of \underline{ultra} high energy cosmic rays. 

\vskip 0.5\baselineskip

\selectlanguage{francais}
\noindent{\bf R\'esum\'e}
\vskip 0.5\baselineskip
\noindent
{\bf Origine des rayons cosmiques  de très haute et ultra haute énergie. }
Alors qu'il existe un certain niveau de consensus sur l'origine Galactique des rayons cosmiques jusqu'au genou  ($E_{k}\sim 3\times 10^{15}$ eV)  et sur leur origine extragalactique au delà de  $\sim 10^{19}$ eV, le débat sur la genèse de ces rayons dans la région intermédiaire a reçu beaucoup moins d'attention, en particulier du fait de l'ambiguïté de la définition même de cette zone. 
L'intervalle d'énergie  de $10^{17}$ eV à $\sim 10^{19}$ eV est probablement celui où la transition Galactique-extragalactique a lieu.
Par conséquent l'origine des rayons cosmiques dans cet intervalle, bien que revêtant une importance toute particulière du point de vu de la physique, est aussi particulièrement difficile à étudier. J'illustre ici quelques idées concernant les sites d'accélération de ces particules et les questions auxquelles leur étude peut répondre, y compris concernant l'origine des rayons cosmique \underline{d'ultra} haute énergie. 

 \vskip 0.5\baselineskip
\noindent{\small{\it Mots-cl\'es~:} VHECR~; UHECR~; genoux~; accélération~; Galactique-extragalactique.}}
\end{abstract}
\end{frontmatter}


\selectlanguage{english}
\section{Introduction}
\label{sec:intro}

The bulk of cosmic rays (CRs) reaching the Earth is most likely accelerated in Galactic sources, and based on energetic arguments, supernova remnants (SNRs) are the most likely sources (see \cite{BlasiRev} for a review of the positive and critical aspects of this scenario). The greatest challenge to this so-called SNR paradigm is represented by the maximum energy that can be achieved: even in the presence of efficient magnetic field amplification at the SNR shock, reaching proton energies in the $\sim PeV$ range appears to be very challenging, so much so that new avenues are being pursued. For instance the perspective of acceleration in the early (fastest) phases of the supernova explosion is insistingly been investigated \cite{schure1,schure2,schure3}, though this version of the SNR paradigm is also not problem free. Many arguments conspire to suggest that SNRs can hardly accelerate CRs to rigidities in excess of $1-10$ PeV. This would imply that the proton spectrum extends to about the knee region, while heavier nuclei with electric charge $Z e$ get accelerated to energies $Z$ times larger. This qualitative picture seems to be confirmed by the fact that the chemical composition at the knee shows a transition from light to heavy \cite{2006JPhCS4741H}. In this scenario it would be natural to expect that Galactic CRs would end with an iron-dominated composition at $\sim 10^{17}$ eV. Yet no appreciable drop in the CR flux is observed at such energies, indicating that either there are Galactic sources able to accelerate to much higher energies than we are able to account for at present, or that there is a substantial flux of extragalactic CRs already at $E\gtrsim 10^{17}$ eV. These two possibilities have very different implications in terms of chemical composition and anisotropy of very high energy (VHE) and ultra high energy (UHE) CRs (see \cite{olinto} for a recent review on UHECRs).  

An extra component of CRs accelerated in the Galaxy was already advocated in \cite{hillas05}, where the extragalactic CR flux was assumed to follow the dip model \cite{dip1} (pure protons). Both possibilities of a Galactic and extragalactic additional component were investigated in \cite{UHEnuclei} in the case in which the extragalactic CR flux has a mixed composition tuned to fit the Pierre Auger Observatory (Auger) observed spectrum and chemical composition. Indeed, the spectra and chemical composition of UHECRs (with energy $\gtrsim 10^{18}$ eV) as measured by Auger \cite{PAOspec,PAOcomp} suggest that CRs are injected with a mixed composition and, somewhat surprisingly, with a rather hard spectrum. 

At energies between the knee and $\sim 10^{18}$ eV the measurement of spectrum and chemical composition is of the utmost importance to understand the origin of VHECRs. Recent measurements carried out with KASCADE-GRANDE \cite{kg1,kg2} reveal an interesting structure in the spectrum and composition of CRs between $10^{16}$ and $10^{18}$ eV: the collaboration managed to separate the showers in electron-rich (a proxy for light chemical composition) and electron-poor (a proxy for heavy composition) showers and showed that the light component (presumably protons and He, with some contamination from CNO) has an ankle like structure at $10^{17}$ eV. The authors suggest that this feature signals the transition from Galactic to extragalactic CRs. The spectrum of Fe-like CRs continues up to energies of $\sim 10^{18}$ eV, where the flux of Fe and the flux of light nuclei are comparable. Similar results were recently put forward by the ICETOP collaboration \cite{icetop}. This finding does not seem in obvious agreement with the results of the Pierre Auger Observatory \cite{PAOcomp}, HiRes \cite{HiResComp1,HiResComp2} and Telescope Array \cite{TAcomp1,TAcomp2}, which show a chemical composition at $10^{18}$ eV that is dominated by the light component. 

Given the complexity of the situation as currently suggested by data, here we will keep the discussion as general as possible. In \S \ref{sec:transition} we will summarize the arguments that lead to require a Galactic or an extragalactic extra CR component; in \S \ref{sec:VHECR} we will discuss some sources of Galactic CRs that may be able to accelerate protons to energies of $\sim 10^{17}$ eV, with special attention for pulsars and powerful supernova explosions. In \S \ref{sec:uhecr} we will extend the discussion on particle acceleration to extragalactic sources and their potential as sources of UHECRs. We will summarize in \S \ref{sec:summary}.

\section{Galactic versus extragalactic CRs: where is the transition?}
\label{sec:transition}

\begin{figure}[t]
  \centering
  \includegraphics[width=0.45\textwidth]{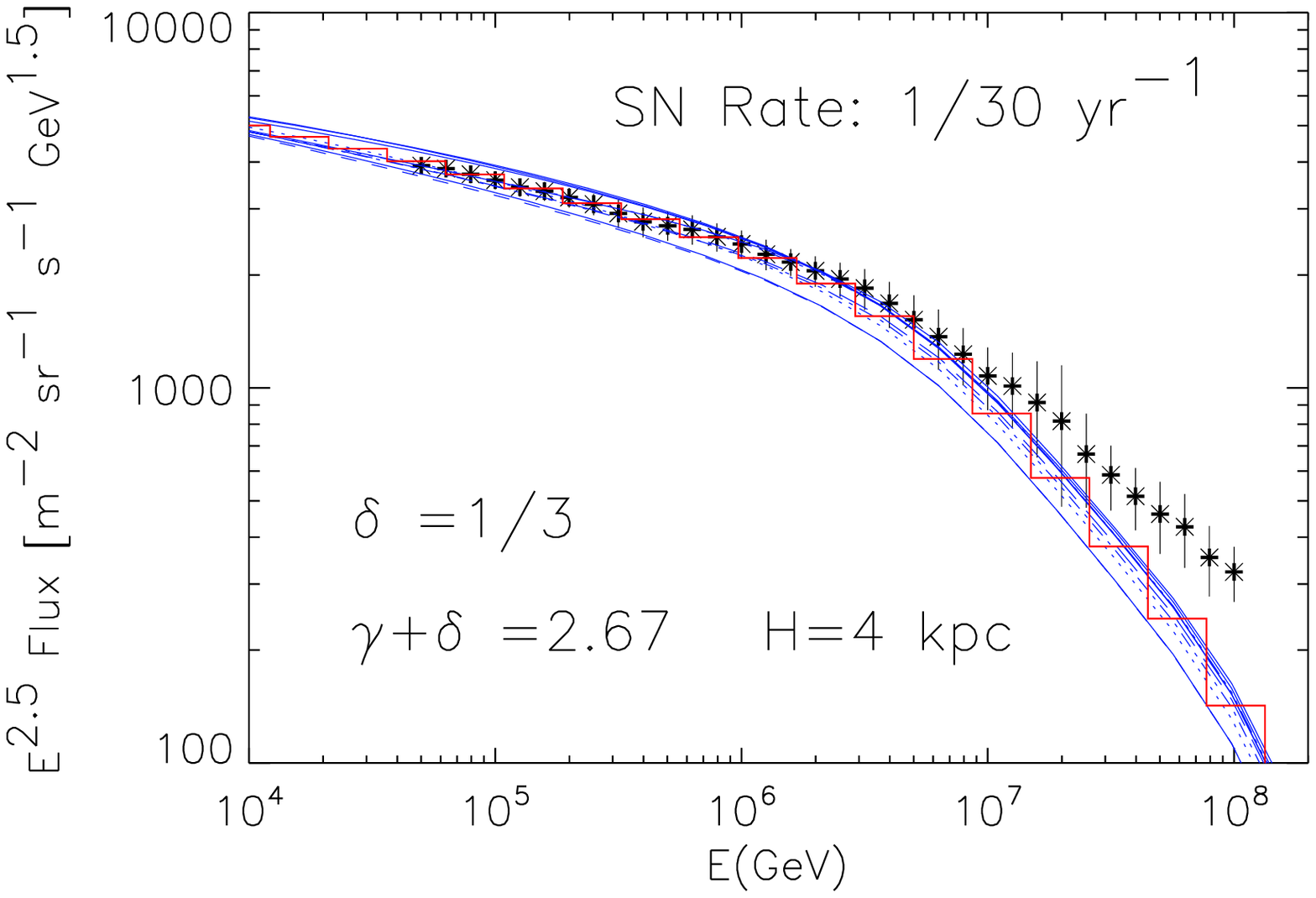}
    \includegraphics[width=0.45\textwidth]{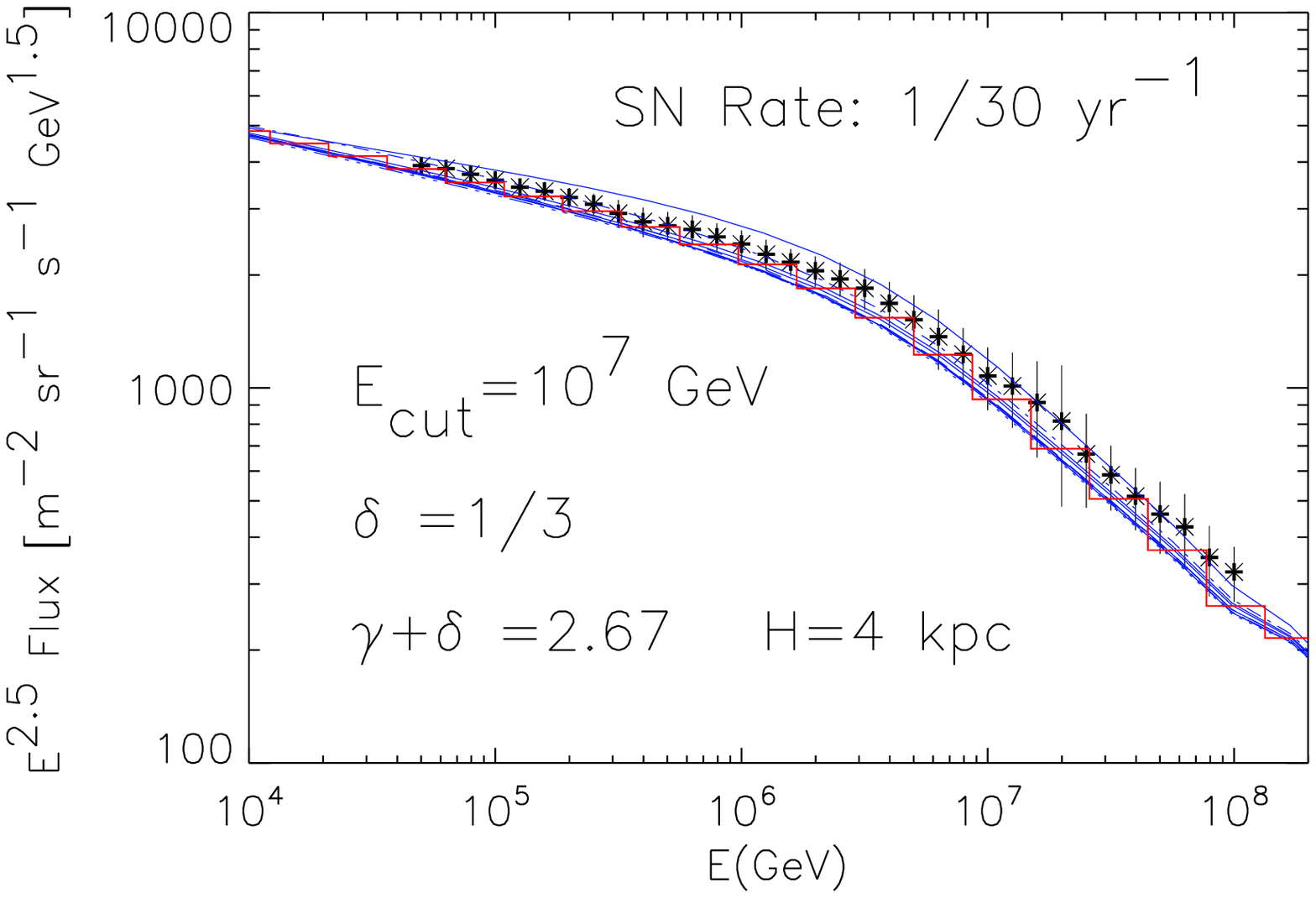}
  \caption{{\it Left Panel:} Spectrum of galactic CRs from SNRs exploded at random in the Galaxy at a rate of one every 30 years. The lines represent ten realizations of the spatial distribution of SNRs, while the red step function shows the average among the ten realizations. The data points are from Ref. \cite{meanspec}. The diffusion coefficient scales as $D(E)\propto E^{\delta}$ with $\delta=1/3$. The injection spectrum is a power law with slope $\gamma$ chosen so that $\gamma+\delta=2.67$. The size of the Galactic halo has been chosen to be $H=4$ kpc. The knee is produced by the overlap of spectra of CR nuclei with rigidity dependent cutoffs \cite{amato1}. {\it Right Panel:} Same as the left panel plus the flux of UHECR protons modeled according to the dip model \cite{dip1} with a low energy cutoff at $10^{7}$ GeV.}
  \label{fig:spec}
 \end{figure}

A reasonable understanding of the CR spectrum up to energies of order $10^{16}$ eV can be achieved by assuming that sources inject a power law spectrum into the interstellar medium (ISM) and that diffusion adds an energy dependence, which reflects in the high energy behaviour of the secondary to primary ratios, such as B/C. One such description can be found in \cite{amato1}, where the all-particle spectrum is fitted reasonably well with a diffusion coefficient $D(E)\propto E^{1/3}$, qualitatively compatible with the observed anisotropy \cite{amato2}. The maximum energy of protons was assumed to be $\sim 5\times 10^{15}$ eV.  This situation is illustrated in the left panel of Fig. \ref{fig:spec}: the different lines show ten different realizations of the spatial distribution of SNRs in the Galaxy, while the step-function shows the average flux of the ten realizations. The data points with error bars represent the average all-particle spectrum as provided in Ref. \cite{meanspec}. The spread in the theoretical predictions provides an estimate of the role of fluctuations on the all-particle spectrum. The high energy part of the all-particle spectrum is dominated by the iron component, with a spectrum that dives starting at $\sim 10^{16}$ eV. Fig. \ref{fig:spec} (left panel) shows the predicted all-particle spectrum departing from the observed one at energies above $\sim 10^{16}$ eV.

At energies $\gtrsim 10^{16}$ eV, a substantial contribution to the all-particle spectrum must come either from an additional class of galactic sources or from extragalactic CRs. For instance in the right panel of Fig. \ref{fig:spec} we plot the all-particle spectrum obtained by summing the SNR contribution (again, ten realizations of their distribution are shown) and a basic dip model \cite{dip1} for the extragalactic component. A low energy exponential cutoff in the extragalactic CR flux has been imposed by hand at $10^{7}$ GeV, in order to mimic the possible effect of extragalactic magnetic fields. The normalization of the predicted extragalactic CR flux was chosen so as to fit the all-particle spectrum as measured by HiRes \cite{hires}. 

The perspective provided by Auger on UHECRs is rather different from that of HiRes and more recently of Telescope Array: the mean depth of shower maximum, $X_{max}(E)$, and its dispersion $\sigma(E)$ as measured by Auger suggest that at energies $\gtrsim 10^{18}$ eV the chemical composition from proton dominated becomes increasingly more dominated by heavy nuclei, a finding that appears to be incompatible with the predictions of the dip model \cite{dip2,dip3}. As a consequence, the issue of the origin of the CRs that fill the gap shown in the left panel of Fig. \ref{fig:spec} might need to be reconsidered. The issue was studied in detail in \cite{UHEnuclei} (see also \cite{taylor}): the $X_{max}(E)$ and $\sigma(E)$ measured by Auger imply that a mixed composition is injected by extragalactic sources, with a hard spectrum with slope $\sim 1-1.6$, at odds with the most common acceleration mechanisms, that lead to steeper spectra (see \S \ref{sec:uhecr} for more discussion of this point). 

The most striking consequence of the hardness of the injection spectra is however that only the highest energy part of the spectrum can be fitted ($E\gtrsim 5\times 10^{18}$ eV) leaving another gap in the spectrum at lower energies: the conclusion as derived in \cite{UHEnuclei} is that an additional component is needed to fit the all-particle spectrum, and such a component must be mostly light in order to accomodate the light composition measured by HiRes \cite{HiResComp1,HiResComp2}, TA \cite{TAcomp1,TAcomp2} and Auger\cite{PAOcomp}) at $10^{18}$ eV. Moreover, this extra component is required to have a steep spectrum with slope $\sim 2.7-2.8$ and a relatively low maximum energy, $\sim 5\times 10^{18}$ eV, otherwise the fit to $X_{max}(E)$ and $\sigma(E)$ at high energies would be negatively affected. 

\begin{figure}[t]
  \centering
  \includegraphics[width=0.6\textwidth]{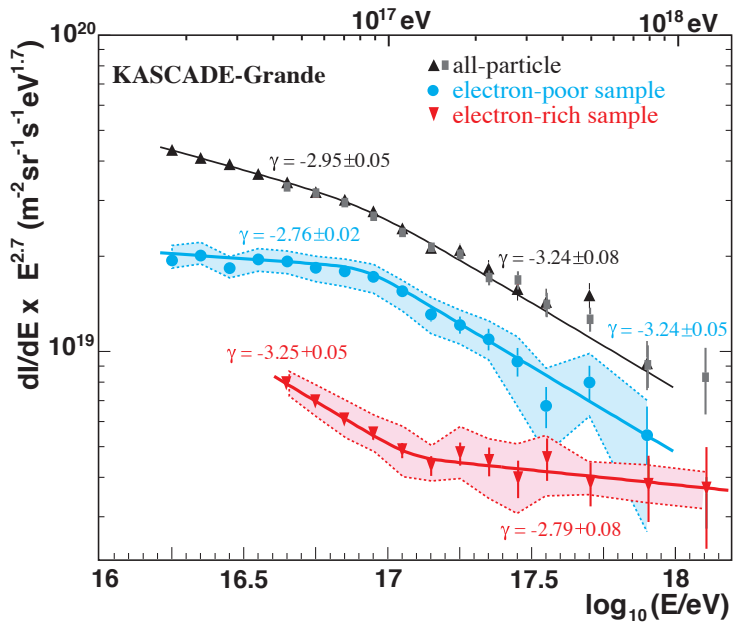}
  \caption{Spectrum of CRs separated in a light component (electron rich showers) and heavy component (electron poor showers) as measured by the KASCADE-Grande experiment \cite{kg2}. The all-particle spectrum is also shown. The lines are the result of a fit to the experimental data as computed in \cite{kg2}.}
  \label{fig:kg}
 \end{figure}

The recent KASCADE-Grande measurement of the spectrum and chemical composition in the energy region between $10^{16}$ eV and $10^{18}$ eV (see Fig. \ref{fig:kg}, from Ref. \cite{kg2}) has shed an interesting light on the transition region: on one hand these measurements show clear evidence of a feature in the light component in the form of an ankle-like structure at $\sim 10^{17}$ eV. This spectral feature is interpreted by the collaboration as the possible signature of a transition between a Galactic component (extending to $\sim 10^{17}$ eV) and an extragalactic light component (with slope $2.79\pm 0.08$). On the other hand, the same measurements show that the spectrum of the heavy CR component has a knee at $\sim 10^{17}$ eV, where it steepens from a slope $2.95\pm 0.05$ to a slope $3.24\pm 0.08$). In \cite{UHEnuclei} it was shown that the spectrum of the heavy component can be interpreted as the result of the superposition of the standard Fe spectrum from SNRs (with a cutoff at $\sim 10^{17}$ eV, 26 times above the knee) and an additional Fe component, with the same spectrum but a cutoff at $10^{18}$ eV.

This scenario shows that the transition region may be rather complicated, with Galactic CR sources contributing heavy ions up to $\sim 1$ EeV and multiple components of extragalactic CRs, with protons and nuclei injected with different injection spectra. Though not appealing, this complexity may be the way in which Nature realizes the transition from galactic to extragalactic CRs, and it has its own interesting aspects: in fact it has the noticeable implication that, contrary to the simplest intuition, there may be Galactic sources able to accelerate CRs to a rigidity of $\sim (3-4)\times 10^{16}$ V (corresponding to iron nuclei at $\sim 10^{18}$ eV), about an order of magnitude above the rigidity of the knee. This exacerbates the severe difficulties faced by the theory of particle acceleration in SNRs, and might require a different class of sources altogether.


In the next two sections we will briefly discuss some issues related to the acceleration of Galactic and extragalactic CRs to energies in excess of $10^{17}$ eV, having in mind the scenario described above. 

\section{Galactic VHECRs}
\label{sec:VHECR}

In this section I focus on the following question: is there a class of Galactic sources that can potentially contribute CR iron nuclei up to energy $E_{max}\approx 10^{18}$ eV? Below I discuss two classes of astrophysical sources as candidate accelerators of VHECRs.

\subsection{Supernova remnants}
\label{sec:snr}

The first temptation is to check whether some peculiar type of SNRs may serve the purpose. For instance, in Ref. \cite{zira2010} it was proposed that supernovae of type IIb, that exploded in the dense wind of the presupernova red giant star, could accelerate to $E_{max}$ of this order of magnitude. This conclusion followed from assuming that all SNRs reach a downstream magnetic field pressure that, independently of the shock velocity ($v_{s}$), is $\sim 3.5\%$ of the ram pressure $\rho v_{s}^{2}$, as a result of streaming instability induced by accelerated particles. This assumption is based on the finding that in the cases in which magnetic field amplification has been inferred from the morphology of X-ray emission, this estimate seems correct as an order of magnitude \cite{volk}. However, the condition for the streaming instability to be effective is that it grows fast enough. Resonant streaming instability does not satisfy this constraint because in the relevant regime its growth is slower than naively thought (see \cite{amato09} and the discussion in \cite{BlasiRev}). This leads to expecting the resonant branch to saturate at $\delta B\sim B_{0}$, where $B_{0}$ is the preexisting magnetic field strength. This has led to several suggestions as to which type of instability could be responsible for the magnetic field amplification. The fastest growing instability is the one discussed in \cite{Bell2004}, which takes place whenever the shock accelerates CRs effectively, namely when $\frac{n_{CR}}{n_{i}}>\frac{v_{A}}{v_{s}}\frac{v_{A}}{c}$, where $v_{A}$ is the Alfv\'en velocity in the unperturbed magnetic field. In terms of the acceleration efficiency this condition can be rewritten as:
\be
\xi_{CR} \gg \frac{\Lambda}{3} \left( \frac{v_{A}}{v_{s}}\right)^{2}\frac{c}{v_{s}}\approx 1.2\times 10^{-3} \left( \frac{v_{s}}{5\times 10^{3} km/s}\right)^{-3},
\label{eq:condition1}
\ee
where we assumed that the spectrum of accelerated particles is a power law $E^{-2}$ up to a maximum energy $E_{max}$ such that $\Lambda=\ln(E_{max}/m_{p}c^{2})\sim 15$ (corresponding to $E_{max}=3\times 10^{6}$ GeV). From Eq. \ref{eq:condition1} one can see that for a typical acceleration efficiency $\xi_{CR}\sim 10\%$, a shock faster than $\sim 1000$ km/s is always in this strongly CR driven regime. Hence the excited waves \cite{Bell2004} are quasi-stationary modes with a growth rate that is fastest on a scale $1/k_{M}\ll r_{L}$ where $r_{L}$ is the Larmor radius of the particles (calculated in the unperturbed magnetic field) that generate the CR current that induces the instability. An estimate of $k_{M}$ is found from the condition of balance between the magnetic tension and the force induced on the plasma by the return current:
\be
k_{M} B_{0} = \frac{4\pi}{c} j_{CR}.
\ee

The physical concept that is becoming increasingly more clear is that the magnetic turbulence necessary for particle scattering is generated by particles with energy $E$ that try to escape the system: their current induces the streaming instability that eventually leads to the creation of the waves that are able to scatter particles with the same energy $E$ at a later time. These particles will therefore return to the shock surface and the process will repeat itself. It can be shown that the relevant CR current is $j_{CR}=n_{CR} \left( \frac{m_{p}c^{2}}{E}\right) e v_{s}$, where $E$ is the energy of escaping particles, while $n_{CR}$ is the total number density of CR particles at the shock surface, with energy above $m_{p}c^{2}$ (in principle the escaping particles leave the acceleration region at speed close to the speed of light, but the current can be written as a function of the shock speed and the CR density at the shock). 

The fastest growing mode, of wavenumber $k_{M}$, has a growth rate
\be
\gamma_{max} = k_{M} v_{A} = \frac{4\pi}{c} \frac{n_{CR} \left( \frac{m_{p}c^{2}}{E}\right) e v_{s}}{\sqrt{4\pi \rho}}.
\ee
It is worth noticing that 
\be
k_{M} r_{L}(E) = \frac{\xi_{CR}}{\Lambda} \left( \frac{v_{s}}{v_{A}}\right)^{2} \left( \frac{v_{s}}{c}\right) \gg 1.
\label{eq:nores}
\ee
Eq. \ref{eq:nores} clearly illustrates the fact that the waves generated by the non resonant streaming instability are excited on spatial scales which are much smaller that the gyration radius and therefore have negligible effects on the scattering of particles with energy $E$. However, it has been advocated that the saturation of this instability, that occurs after a few turnarounds, say $\gamma_{max} \tau \sim 5$, also leads to a sort of inverse cascading, namely to the creation of power on scales much larger than $1/k_{M}$. In fact one can make a case that the instability stops when perturbations on a scale $\sim r_{L}^{*}$ have been generated. Here $r_{L}^{*}$ indicates the Larmor radius in the amplified magnetic field, that by the time the saturation has been reached is much larger than the preexisting magnetic field. 

Imposing the condition $\gamma_{max} \tau\sim 5$ for the case of a SNR in a uniform background medium of density $\rho$ one easily obtains the maximum energy to be \cite{schure1,schure2,schure3}:
\be
E_{M} = \frac{\xi_{CR}}{10\Lambda} \frac{\sqrt{4\pi \rho}}{c} e R v_{s}^{2},
\ee
where $R$ is the radius of the blast wave. As a numerical value we can assume the radius $R$ to coincide with the beginning of the Sedov phase, when an amount of mass equal to the mass of the ejecta has been processed, $R\sim 2 ~ M_{ej,\odot}^{1/3} n_{ISM}^{1/3}~ pc$, and the interstellar medium density $n_{ISM}$ is normalized to 1 $cm^{-3}$. The shock velocity is normalized by using the total energy of the SN explosion, $(1/2) M_{ej} v_{s}^{2} = E_{SN}$, where $M_{ej}$ is the mass of ejecta. It follows that 
\be
E_{M} = \frac{\xi_{CR} 3^{1/3}}{5 \Lambda} \frac{e}{c} \left( 4\pi \rho \right)^{1/6} E_{SN} M_{ej}^{-2/3} \approx 2\times 10^{5}~\rm GeV,
\ee
where the numerical value has been obtained by assuming that $M_{ej}=1~M_{\odot}$, $n_{ISM}=1~cm^{-3}$ and $E_{SN}=10^{51}$ erg. As discussed in Ref. \cite{schure2}, a typical SN exploding in the ISM has a maximum energy that is likely in the range of a few hundred TeV, falling short of the knee by about one order of magnitude. 

The argument presented above can also be generalized to the case of a supernova explosion that takes place in the red supergiant wind of the parent star: the density profile in this case can be written as
\be
\rho(r) = \frac{\dot M}{4\pi r^{2} v_{W}},
\ee
where $\dot M$ is the rate of mass loss of the star, typically around $10^{-5} M_{\odot} yr^{-1}$ and $v_{W}\sim 10$ km/s is the wind velocity. The magnetic field in the wind, before the passage of the SN blast wave is usually assumed to be in equipartition with the kinetic energy of the wind:
\be
\frac{B_{0}^{2}(r)}{4 \pi} = \rho(r)  v_{W}^{2} \to B_{0}(r) = \left(\dot M v_{W}\right)^{1/2} \frac{1}{r}.
\ee
In the case of expansion of the blast wave in the wind of the progenitor, the argument on the saturation of the current driven instability leads to 
\be
E_{M}(R) = \frac{\xi_{CR} e}{5\Lambda c} \sqrt{\frac{\dot M}{v_{W}}} v_{s}^{2}(R).
\label{eq:wind}
\ee
For the case of expansion in the wind, the Sedov phase starts at a distance $R=M_{ej} v_{W} / \dot M$ from the location of the explosion and the shock velocity scales with time as $v_{s} \sim t^{-\frac{1}{m-2}}$, where $m$ is the parameter that describes the shape of the density profile of the ejecta ($\rho_{ej}\propto r^{-m}$). The parameter $m$ is usually assumed to be $m=9$ for core collapse supernovae and $m=7$ for type Ia supernovae. The value $m=9$ leads to $v_{s}\sim t^{-1/7}$, namely the velocity drop is rather slow during the ejecta dominated phase of the expansion, that for typical values $\dot M = 10^{-5} M_{\odot} yr^{-1}$, $v_{W}=10$ km/s and $M_{ej}=1 M_{\odot}$ lasts for about 50 years for ejecta velocity of order $20000$ km/s (corresponding to a total energy $E_{SN}\sim 4\times 10^{51}$ erg). Using these values in Eq. \ref{eq:wind}, we get a maximum energy $E_{M}\approx 2 \times 10^{6}$ GeV, close to the position of the knee. Given the time dependence of the shock velocity, one can envision that the maximum energy may actually grow to larger values at earlier times during the ejecta dominated phase, although the total energy in accelerated particles is bound to be small because of the limited amount of mass processed by the shock at those times. Nevertheless we cannot exclude that a sizeable contribution to the flux of CRs with energy above the knee may come from these very early phases of the supernova explosion. It is clear however that all parameters need to be pushed to their extreme values in order to realize this situation: for instance one could speculate about higher acceleration efficiency, $\xi_{CR}=0.2$, higher mass loss rate, $\dot M=10^{-4} M_{\odot} yr^{-1}$ and somewhat higher values of the shock velocity, $30000$ km/s, so as to have $E_{M}\sim 2\times 10^{7}$ GeV. 

In Ref. \cite{zira2010} the authors argued that SN IIb ($E_{SN}\sim 3\times 10^{51}$ erg, $M_{ej}=1 M_{\odot}$, $\dot M = 10^{-4} M_{\odot} yr^{-1}$) can reach a proton maximum energy around $\sim 5\times 10^{7}$ GeV. For this choice of parameters, their strong assumption that the magnetic energy density downstream of the shock is always $3.5\%$ of the ram pressure of the plasma is not too severe, in that the estimates described above lead to a magnetic pressure which is only a factor $\sim 3$ lower than this benchmark value. Moreover, \cite{zira2010} quote a rate of SN IIb explosions that is only $\sim 10$ times lower than more frequent supernovae, partially compensated by the larger energetics (about three times larger). 

Several aspects of the problem of particle acceleration should be kept in mind here: the geometry of the background magnetic field in the wind of the presupernova star is all but trivial. If to make a parallel with the case of the solar wind, one would expect that the field geometry is mostly perpendicular to the 
shock normal, and this may affect the considerations presented above in at least two ways. First, the rate of growth of the waves is lower than estimated above, being maximal for the case of parallel configuration. Second, acceleration might proceed in the regime of perpendicular transport \cite{joki87}, namely with a somewhat higher rate of acceleration. The first effect implies a lesser role of CR induced instabilities upstream of the shock, while the second effect might help in reaching somewhat higher energies \cite{joki87,giaca1,giaca2}.

\subsection{Pulsars and their nebulae}
\label{sec:ns}

A rotating magnetized neutron star generates an induced electric field. The electric properties of the neutron star can be expressed in terms of quantities measured at the light cylinder, located at a radius $R_{L}$ such that $\Omega R_{L}=c$, where $\Omega=2\pi/P$ is the angular frequency of rotation of a pulsar with period $P$. In general the rotation axis and the magnetic axis do not coincide, but for the purpose of this discussion we will ignore this complication, since we are interested in a qualitative description of the acceleration process. In principle, the maximum voltage that is accessible to particles can be expressed as \cite{bible}:
\be
\phi = \frac{B_{s}\Omega^{2} R_{s}^{3}}{2c^{2}} .
\ee
If a charge $Ze$ could make use of this voltage drop, its maximum energy would be 
\be
E_{max} = Z e \phi = 1.5 Z\times 10^{19} eV \left( \frac{B_{s}}{10^{13} G} \right)
 \left( \frac{\Omega}{3000 s^{-1}} \right)^{2}  \left( \frac{R_{s}}{10^{6} cm} \right)^{3}.
\label{eq:maxE}
\ee
In general, several factors may limit the actual maximum energy of hadrons to much lower values, especially because of energy losses. Inside the light cylinder, the main channel of energy losses for nuclei is represented by the emission of curvature radiation which leads to a rate of energy loss: 
\be
\frac{dE}{dt} = - \frac{2}{3c} \Gamma^{4} e^{2} \Omega^{2} ,
\ee
where we assumed that the curvature of the field lines is of the order of the radius of the light cylinder $R_{L}$. If we impose that the rate of energy gain and losses of a nucleus of charge $Ze$ are equal at the maximum energy for a gap size $\xi R_{L}$:
\be
\Gamma_{max} = \left( \frac{3}{2}\frac{B_{s}R_{s}^{3}\xi^{-3} \Omega}{Ze c}\right)^{1/4} = 2.4\times 10^{8} \xi^{-3/4}Z^{-1/4} \left( \frac{B_{s}}{10^{13} G} \right)^{1/4} 
 \left( \frac{\Omega}{3000 s^{-1}} \right)^{1/4}  \left( \frac{R_{s}}{10^{6} cm} \right)^{3/4}.
\label{eq:Gmax}
\ee
For the reference values of the parameters and $\xi\sim 1$, an iron nucleus could reach energies as high as $(1-5) \times 10^{18}$ eV, well below the maximum energy that could be estimated based on the whole available potential drop, but still large enough to be of relevance for the problem of the continuation of the CR spectrum above the knee, as discussed above. On the other hand, other issues arise that we have not discussed yet: the electric force on both electrons and nuclei at the neutron star surface exceeds by many orders of magnitude the gravitational force on the same particles; electrons freely flowing in the lattice that makes the surface of the star can be easily extracted and injected in the magnetosphere. In fact their curvature radiation leads to production of photons that are still above threshold for pair production in the strong neutron star magnetic field, so that an electromagnetic cascade develops. For each electron extracted from the surface, $10-10^{6}$ secondary electrons and positrons may be produced (the so-called multiplicity \cite{multiplicity}). These pairs fuel a relativistic wind that is launched from the light cylinder region. The charge density is bound to stay close to the Goldreich-Julian density \cite{goldreich}. For nuclei the situation is quite different: the surface of the neutron star is likely to be made of heavy nuclei, probably iron, although some fall-back material from the supernova explosion may be present (hydrogen and helium). The possibility to extract nuclei from the lattice depends on poorly known details of the structure of the surface, but it appears to be possible at least  in principle \cite{bible}. The density of nuclei in the magnetosphere is however bound to be close to the Goldreich-Julian density (nuclei are not subject to cascading). Since the mass of the nuclei is $A m_{p}/m_{e}$ larger than the electron mass, even a small fraction of nuclei in the wind may have important dynamical effects. 

The wind is terminated at a shock where particle acceleration occurs, as shown by observations of radio, X and gamma ray emission from the postshock region in known pulsar wind nebulae (see \cite{slane} for a review). It is in fact the emission from particles accelerated at such termination shock that extends to higher energies, rather than the emission from the particles directly accelerated at the pulsar. In any case an upper bound to the maximum energy is the total potential drop available because of the rotation of the magnetized star. As discussed above, this may in principle lead to acceleration of iron nuclei up to $\sim 4\times 10^{20}$ eV for very young, rapidly spinning neutron stars \cite{epstein}.

The spectrum of CRs contributed by neutron stars is highly uncertain: in principle the spectrum of iron nuclei accelerated by the electric field in the magnetosphere is peaked at $A m_{p} \Gamma_{max}$, with $\Gamma_{max}$ given by Eq. \ref{eq:Gmax}, which is weakly dependent upon $\Omega$ and therefore weakly dependent upon time since $\Omega$ decreases because of the spin down, $\dot \Omega \propto \Omega^{n}$, with $n$ the breaking index. In fact the final Lorentz factor of nuclei could be higher than $\Gamma_{max}$ if they get advected with a faster wind. As discussed in \S \ref{sec:uhecr}, this possibility depends on the pair multiplicity, and can be realized for sufficiently small multiplicities. 

These simple estimates show that at least in principle young pulsars can contribute a flux of CR nuclei up to energy $E\sim 10^{18}$ eV,  even in the presence of curvature energy losses. The spectrum of these CRs is unknown: in principle, as long as the Lorentz factor of accelerated nuclei is determined by either curvature losses or the wind dynamics, one may expect that at any given time the maximum Lorentz factor is peaked around a given value and that a larger spread in energy may derive from integration over the history of the pulsar spin down. On the other hand, some level of acceleration can also take place at the termination shock.

In the next section we will also discuss an extension of this picture to ultra high energy particles, as was originally presented in \cite{epstein,arons} and later considered in more detail in \cite{FangOlinto1,FangOlinto2}.

\section{Ultra high Energy Cosmic Rays}
\label{sec:uhecr}

The data on spectrum and mass composition collected with Auger have affected enormously our perspective on the problem of the origin of ultra high energy cosmic rays: a decade ago the common wisdom was that UHECRs were most likely protons accelerated to maximum energies $\gtrsim 10^{20}$ eV and that the detection of the sources of these very energetic particles was behind the corner, since no Galactic or intergalactic magnetic field could possibly deflect protons at $10^{20}$ eV by more than a few degrees. Indeed, detailed simulations of the number of events necessary to detect the sources, for different assumptions on the physical parameters of the sources, were carried out in Refs. \cite{demarco1,demarco2,demarco3} for both the expected statistics of Auger and EUSO. 

Things have changed quite a bit: current Auger data are consistent with a light mass composition at energies around $\sim 10^{18}$ eV, with a gradual transition to a heavier mass composition at higher energies. This indication comes from both the mean penetration depth of showers, $\langle X_{max}(E)\rangle$, and its dispersion, $\sigma(E)$: the former quantity decreases with increasing mass of the primary nucleus since heavier particles interact higher in the atmosphere, corresponding to a small traversed grammage. The dispersion around the mean $X_{max}$ is also smaller for heavier nuclei since the larger number of nucleons in the nucleus leads to smaller fluctuations in the shower development. Both these trends are observed by Auger \cite{PAOcomp}. In fact a light mass composition around $10^{18}$ eV is also observed by the other two large CR detectors, HiRes \cite{HiResComp1,HiResComp2} and Telescope Array (TA) \cite{TAcomp1,TAcomp2}. The possible discrepancy appears at higher energies where both HiRes and TA data suggest a light composition. It is however important to keep in mind that the level of agreement (or disagreement) between the results of HiRes and TA and those of Auger depend on the particular interaction model used to interpret the values of $\langle X_{max}(E)\rangle$ \cite{agree}. 

The decreasing trend in $\sigma(E)$ has in fact even deeper implications: in general, if at some given energy $E$, proton showers and, say, iron showers were present at comparable rates, the global distribution of penetration depth would not be intermediate between that of protons and iron, but it would rather have a width comparable with or even larger than that of proton induced showers. The fact that $\sigma(E)$ decreases actually implies that at a given energy the lighter components have started to disappear. For instance, the flux of protons at $10^{19}$ eV must have already been dropping, because the dispersion at the same energy is appreciably smaller than for proton induced showers. This simple consideration, together with the proton dominance at $10^{18}$ eV, implies that Auger data require a proton flux with a pronounced suppression at $E_{max}^{p} \sim 5\times 10^{18}$ eV. Since this suppression is not associated with energy losses during propagation, the natural conclusion of this line of thought is that the sources of UHECRs accelerate protons up to $\sim 5 \times 10^{18}$ eV and iron (if present) up to $\sim 10^{20}$ eV. These simple concepts find a clear implementation in the calculations presented in \cite{UHEnuclei,taylor}, where it was also shown that spectrum and mass composition as observed by Auger can be properly described only by assuming that the injection spectrum is very flat, $\propto E^{-\gamma}$, with $\gamma\sim1-1.6$, quite different from the spectra that are usually derived based on diffusive shock acceleration at both non-relativistic and relativistic shocks. This conclusion might be affected however by the presence of magnetic fields, either inside the sources or in the propagation volume: the actual injection spectrum might be cut off at low energy because of propagation effects. 

The calculations in Ref. \cite{UHEnuclei} also show that this unexpected picture seems to provide a description of the data only at energies $\gtrsim 5\times 10^{18}$ eV, while underestimating the flux at lower energies. An additional component is required to avoid this problem: such component is bound to be made of extragalactic protons in order to fulfill the constraints on mass composition and anisotropy at $10^{18}$ eV \cite{PAOani}, and to have a steep spectrum, with slope $\sim 2.7-2.8$ in order to avoid affecting the high energy mass composition, as described above. As discussed in \S \ref{sec:transition}, this component might be the light component recently measured by KASCADE-Grande \cite{kg1,kg2}.


The proliferation of components with different maximum energies and different spectral shapes needed to achieve a self-consistent description of spectra and mass composition in the energy region above $10^{17}$ eV makes it rather hard to speculate upon the sources of these CRs. As discussed in the previous section, the heavy component identified by KASCADE-Grande could well represent the contribution of Galactic sources, for instance supernovae in the early phases of their expansion or young pulsars.

The extragalactic components are however more problematic, though we can dare put forward some general comments: 
\vskip .5 cm
\begin{itemize}
\item[1.]
The extragalactic sources of CRs are only required to accelerate particles with charge $Ze$ to a maximum energy $\sim (5-10)Z\times 10^{18}$ eV, rather than $\gtrsim 10^{20}$ eV, as a result of Auger data on mass composition. The constraints illustrated by \cite{waxman} and summarized by \cite{myrev} are easily satisfied by many classes of sources, which in fact means that we cannot draw strong conclusions at present as to where these CRs are accelerated.

\item[2.]
Extragalactic UHECRs have a mixed composition when they reach the Earth, therefore acceleration must be taking place in regions with a strong metallicity, so as to make heavy nuclei readily available for acceleration, although some acceleration processes have intrinsic preference for highly charged nuclei (diffusive shock acceleration is one of these). Based on this consideration, some scenarios may be considered as disfavored. For instance, it appears unlikely that UHECR nuclei may originate in shocks formed during large scale structure formation, since these shocks propagate in regions that are not expected to have been polluted by metals as yet. Shocks propagating in the intracluster volume (for instance during cluster mergers) may process metal enriched material, but they are usually rather weak, with typical Mach number $M_{s}\sim \sqrt 2-2$ \cite{gabici}. On the other hand, sources of cosmic rays located inside clusters of galaxies can actually produce nuclei and give rise to interesting phenomenology as due to the presence of the intracluster magnetic field (for instance see Ref. \cite{kotera}).

\item[3.]
The bulk of UHECRs appear to have a hard injection spectrum, with slope $\sim 1-1.6$. This conclusion is at odds with typical acceleration mechanisms as we know them: for instance diffusive shock acceleration in the non-relativistic regime leads to slope $\sim 2$ while at relativistic shocks the slope is $\sim 2.3$ \cite{rel1,rel2,rel3,rel4}. For relativistic shocks even steeper spectra can be obtained because of particle trapping downstream \cite{lemoine,trapping,trapping1}, which implies low acceleration efficiency. Hard injection spectra can be more easily accommodated in acceleration mechanisms based on direct electric fields, such as reconnection and unipolar induction. A caveat in this line of thought is that the presence of relatively strong magnetic fields in the intergalactic medium (at the $nG$ level) can mimic the effect of hard injection spectra, while the actual injection spectra can be somewhat steeper \cite{antiGZK,mollerach} and the flux of low energy particles is suppressed because the propagation time from the closest source can exceed the age of the universe (the so-called magnetic horizon). It is probably worth keeping in mind that the nG field required for this scenario to be effective is at the level of the equipartition field in voids:
\be
\frac{B_{eq}^{2}}{4\pi} \approx \frac{\Omega_{b} \rho_{cr}}{m_{p}} k T_{voids} \to B_{eq} \approx 4 T_{5}^{1/2}~\rm nG,
\ee
where $T_{5}=T_{voids}/10^{5}K$ is the temperature of voids in units of $10^{5}$ K. Since in general equipartition arises from plasma currents and the latter are the symptom of some level of turbulence, it seems unlikely that voids may be subject to such conditions. Even in clusters of Galaxies, which are virialized structures, the inferred magnetic field is below the equipartition level. Nevertheless, $nG$ magnetic fields are still compatible with existing Faraday rotation measurements \cite{burles}, and as such, their existence cannot be ruled out at present. 

\item[4.]
If at the highest energies the mass composition is indeed dominated by heavy nuclei, the possibility to ``see'' the sources directly, through small scale anisotropies seems hard to realize, since even the Galactic magnetic field alone is sufficient to cause large deflections in the trajectory of iron nuclei. 

\end{itemize}
\vskip .5cm
It is interesting to note that a model for the origin of UHECRs that predicted heavy mass composition at high energy and flat injection spectra was proposed in \cite{epstein}, long before Auger published its data. The model is based on the idea, discussed in \S \ref{sec:ns}, that the electric field induced by the neutron star rotation at the surface of the star could extract iron nuclei from the surface and eventually accelerate them in the so-called gap, where the electric and magnetic field are parallel. The possibility that the relevant pulsars could be extragalactic and that the extracted particles were protons rather than nuclei was later considered in \cite{arons}. More recently \cite{FangOlinto1,FangOlinto2} developed this model considering the acceleration of nuclei in both Galactic and extragalactic pulsars and showed that after taking into account spallation in the parent supernova ejecta expanding outwards and surrounding the very young pulsar, and the photodisintegration of nuclei during propagation, one can achieve a satisfactory description of Auger data. 

As discussed in \S \ref{sec:ns}, the maximum energy of nuclei accelerated by the electric field in the gap is typically limited by curvature losses. At any given time the spectrum is expected to be peaked at the maximum energy, so that the total spectrum injected by the neutron star is the result of integral over time of these peaked spectra and leads to $N(E)\propto E^{(1-n)/2}$ where $n$ is the breaking index of the pulsar \cite{epstein}. For $n=3$ (magnetic dipole), the spectrum is $N(E)\sim E^{-1}$. If curvature radiation were the limiting factor, young pulsars could not accelerate UHECR nuclei. However one should keep in mind that particles (electrons and nuclei) are all advected with the wind that is launched somewhere outside the light cylinder. If the Lorentz factor of the wind (strongly dependent upon the multiplicity of the electron-positron pairs created in the magnetosphere) is higher than the maximum Lorentz factor of nuclei, limited by curvature losses, then nuclei would anyway be carried away at the wind Lorentz factor. More quantitatively, a necessary (but not sufficient) condition for the model to be viable is that the pair multiplicity $\eta$ satisfies the condition $\eta\lesssim \eta_{cr}=2m_{p}/m_{e}\sim 4\times 10^{3}$. Unfortunately, accurate models of the pair multiplicity in newborn neutron stars, relevant for the problem of UHECR production, are all but well developed: in these sources the parent electrons extracted from the surface cross the gap while suffering curvature radiation losses as well as inverse Compton losses in the intense radiation field following the supernova explosion, and it is not clear what happens in this environment. If the multiplicity does not exceed $\eta_{cr}$ it is possible that accelerated nuclei can take away a large fraction of the potential drop available in the neutron star magnetosphere, even if they suffer curvature radiation losses while being closer to the star's surface. 

\section{Summary}
\label{sec:summary}

The recent KASCADE-Grande and Auger data on spectra and mass composition have led us to depict a scenario for the origin of VHECRs that seems very different from the one that we had in mind a few years ago. First, the region of the transition from Galactic to extragalactic CRs appears to be very complex, with Galactic iron reaching up to energies of $\sim 10^{18}$ eV, about one order of magnitude larger than what the SNR paradigm would suggest. In the same transition region a light component (protons and helium nuclei) is needed, based on both the KASCADE-Grande data \cite{kg1,kg2} and Auger data. These particles are bound to have an extragalactic origin in order to avoid the anisotropy bounds imposed by Auger at $10^{18}$ eV \cite{PAOani}, and a steep spectrum with slope $\sim 2.7-2.8$. 

Second, according to Auger data, UHECRs with energy $\gtrsim 10^{18}$ eV are mostly nuclei and protons with very hard injection spectra (slope $\sim 1-1.6$), and a maximum energy $\sim 5 Z\times 10^{18}$ eV. Plenty of extragalactic sources satisfy the requirements to reach such maximum energies, as one can realize by taking a quick look at the so-called Hillas plot (see a revised version of the plot in Ref. \cite{olinto}). On the other hand, the hardness of the required injection spectra is rather unusual and may help restricting the classes of sources responsible for the acceleration of these UHECRs. 

As discussed in \cite{schure1,schure2,schure3}, core collapse supernova explosions occurring in the wind of the red giant parent star may potentially reach maximum energies close to the knee (for protons) in the first 20-30 years after the explosion \cite{schure1,schure2,schure3}, when the Sedov phase starts (the early beginning of the Sedov phase is due to the high density in the wind). Interestingly, even higher energies can be reached at earlier times, when however a smaller fraction of the mass has been processed, so that a correspondingly smaller fraction of the supernova energy can be channelled to accelerated particles. It is not yet clear what is the contribution of these very early phases to the CR spectrum above the knee (especially in terms of nuclei), but in principle powerful supernova shells (expanding at $\gtrsim 30000$ km/s) may reach very high energies in such phases.

When the supernova explosion leaves behind a rotating magnetized neutron star, the electric fields generated in the magnetosphere may be potentially very effective in accelerating particles. For instance, for the parameters of the Crab pulsar, Eq. \ref{eq:maxE} returns a maximum achievable energy for iron nuclei of $\sim 4\times 10^{17}$ eV. Larger energies can be reached in principle at earlier times in the pulsar history. Whether this potential drop is fully accessible to nuclei is however a harder question to answer: curvature energy losses may limit the maximum energy of nuclei to lower values, although what really limits the maximum achievable energy is the electron-positron multiplicity in the pulsar magnetosphere, since this parameter determines the Lorentz factor that can be reached by the relativistic wind of pairs launched by the pulsar, that carries away nuclei (if present) together with pairs. Another important question that remains unanswered is whether the electric field at the neutron star surface is strong enough to extract nuclei from the lattice structure that forms the surface of the star. 

Interestingly, the same mechanism could provide UHECR nuclei from newborn pulsars, as first suggested in \cite{epstein,arons} and later investigated in more detail in \cite{FangOlinto1,FangOlinto2}. Although there are still many aspects of the particle acceleration physics in the pulsar surroundings that need to be understood at a deeper level, this scenario appears to be able to provide at the same time the observed mass composition and the hard spectra required to fit Auger data. 

\section*{Acknowledgements}

I am grateful to E. Amato and R. Aloisio for continuous discussion on the topics illustrated here.

\end{document}